\documentclass[aps,prl,showpacs,
  superscriptaddress,
  floatfix,
  preprint,
  amssymb,
  amsmath]{revtex4}
  
\pdfoutput=1

\usepackage{graphicx}
\usepackage{dcolumn}
\usepackage{amsmath}
\usepackage{amsfonts}
\usepackage{amssymb}
\usepackage{bm}
\usepackage{color}
\usepackage{ulem}
\usepackage{longtable}
\usepackage[linkcolor=blue,citecolor=blue,colorlinks=true,pdfstartview=FitH,pdfauthor=Gepraegs]{hyperref}

\begin{document}

\preprint{\today, F.D. Czeschka {\it et al.}, Version v10}

\title{Giant magnetic anisotropy changes in Sr$_{\mathbf{2}}$CrReO$_{\mathbf{6}}$ thin films on BaTiO$_3$}
 
\author{F.D.~Czeschka}
 \affiliation{Walther-Mei{\ss}ner-Institut, Bayerische Akademie der Wissenschaften,
              D-85748 Garching, Germany}

\author{S.~Gepr\"{a}gs}
 \affiliation{Walther-Mei{\ss}ner-Institut, Bayerische Akademie der Wissenschaften,
              D-85748 Garching, Germany}

\author{M.~Opel}
 \affiliation{Walther-Mei{\ss}ner-Institut, Bayerische Akademie der Wissenschaften,
              D-85748 Garching, Germany}

\author{S.T.B.~Goennenwein}
 \affiliation{Walther-Mei{\ss}ner-Institut, Bayerische Akademie der Wissenschaften,
              D-85748 Garching, Germany}

\author{R.~Gross}
\email{Rudolf.Gross@wmi.badw.de}
 \affiliation{Walther-Mei{\ss}ner-Institut, Bayerische Akademie der Wissenschaften,
              D-85748 Garching, Germany}
 \affiliation{Physik-Department, Technische Universit\"{a}t M\"{u}nchen, D-85748
              Garching, Germany}

\date{\today}

\begin{abstract} % max. 100 words for APL
The integration of ferromagnetic and ferroelectric materials into hybrid heterostructures yields multifunctional systems with improved or novel functionality. We here report on the structural, electronic and magnetic properties of the ferromagnetic double perovskite Sr$_2$CrReO$_6$, grown as epitaxial thin film onto ferroelectric BaTiO$_3$. As a function of temperature, the crystal-structure of BaTiO$_3$ undergoes phase transitions, which induce qualitative changes in the magnetic anisotropy of the ferromagnet. We observe abrupt changes in the coercive field of up to 1.2\,T along with resistance changes of up to 6.5\%. These results are attributed to the high sensitivity of the double perovskites to mechanical deformation.

\end{abstract}


\maketitle

Composite or hybrid material systems consisting of ferromagnetic (FM) and ferroelectric (FE) compounds have attracted increasing interest over the last
years. They provide strong magnetoelectric coupling, making them promising for
new storage devices \cite{Eerenstein2006,Novosad2000,Gepraegs2007}. Moreover,
ferromagnetic thin film/ferroelectric substrate heterostructures allow to
study the magnetic properties of one and the same ferromagnetic layer under
different strain conditions \cite{Lee2000,Brandlmaier2008,Weiler2009,Vaz2009,Tian2008}. In this context,
ferromagnetic materials with a strong magnetocrystalline coupling are very
attractive. Promising materials are ferromagnetic double perovskites
\cite{Kobayashi1998} such as Sr$_2$CrReO$_6$ (SCRO), which show a giant
anisotropic magnetostriction \cite{Serrate2007} caused by a large orbital
moment on the Re site \cite{Majewski2005b}. In addition, SCRO has a Curie
temperature of 635\,K \cite{Kato2002}, well above room temperature, and a
predicted high spin polarization of 86\% \cite{Vaitheeswaran2005}. Here we use
Sr$_2$CrReO$_6$/BaTiO$_3$ (FM/FE) heterostructures to investigate the magnetic and transport
properties of SCRO thin films under different strain conditions, making use of
the structural phase transitions of BaTiO$_3$ (BTO) upon temperature variation
\cite{Kay1949}: Below 393\,K, BTO becomes ferroelectric, and its lattice
structure changes from cubic to tetragonal. Within the ferroelectric state,
the lattice symmetry is further reduced to orthorhombic (below 278\,K), and
finally to rhombohedral (below 183\,K). The dielectric constant,
the spontaneous polarization and the lattice constants change abruptly at these
phase transitions accompanied by a thermal hysteresis \cite{Kay1949}. Our
experiments show a giant change of the magnetic anisotropy and huge coercive
fields, depending on the substrate induced strain state.

The choice of SCRO/BTO as FM/FE heterostructure is obvious as the deposition of epitaxial
SCRO/BTO heterostructures should be straightforward, since the growth of SCRO
films with high crystalline and magnetic quality has been demonstrated on
SrTiO$_3$ (STO) substrates, which have similar crystal structure and lattice
constants as BTO \cite{Asano2004,Gepraegs2009}. The SCRO films used in this study were
grown via pulsed laser deposition \cite{Klein1999} in a pure
oxygen atmosphere of $6.6\times 10^{-4}$\,mbar and a substrate temperature of
700$^\circ$\,C. These parameters were found to be optimal for SCRO
on SrTiO$_3$ \cite{Gepraegs2009}. The films were characterized by high
resolution x-ray diffraction (HRXRD), magneto transport measurements in a
four-point geometry, and SQUID magnetometry. Note that the saturation magnetic
field of SCRO exceeds several Tesla \cite{Michalik2007}, so that a determination
of the diamagnetic (substrate) contribution to the SQUID signal is difficult.
We therefore subtracted a straight line from all $M(H)$-loops shown below such that the slope of the signal for high fields becomes zero.

%%%%%%%%%%%%%%%%% FIGURE 1: XRD %%%%%%%%%%%%%%%%%%%%%%%%%%%%%%%%%%%%%
\begin{figure}[tb]
    \centering
    \includegraphics [width=0.85\columnwidth,clip=]{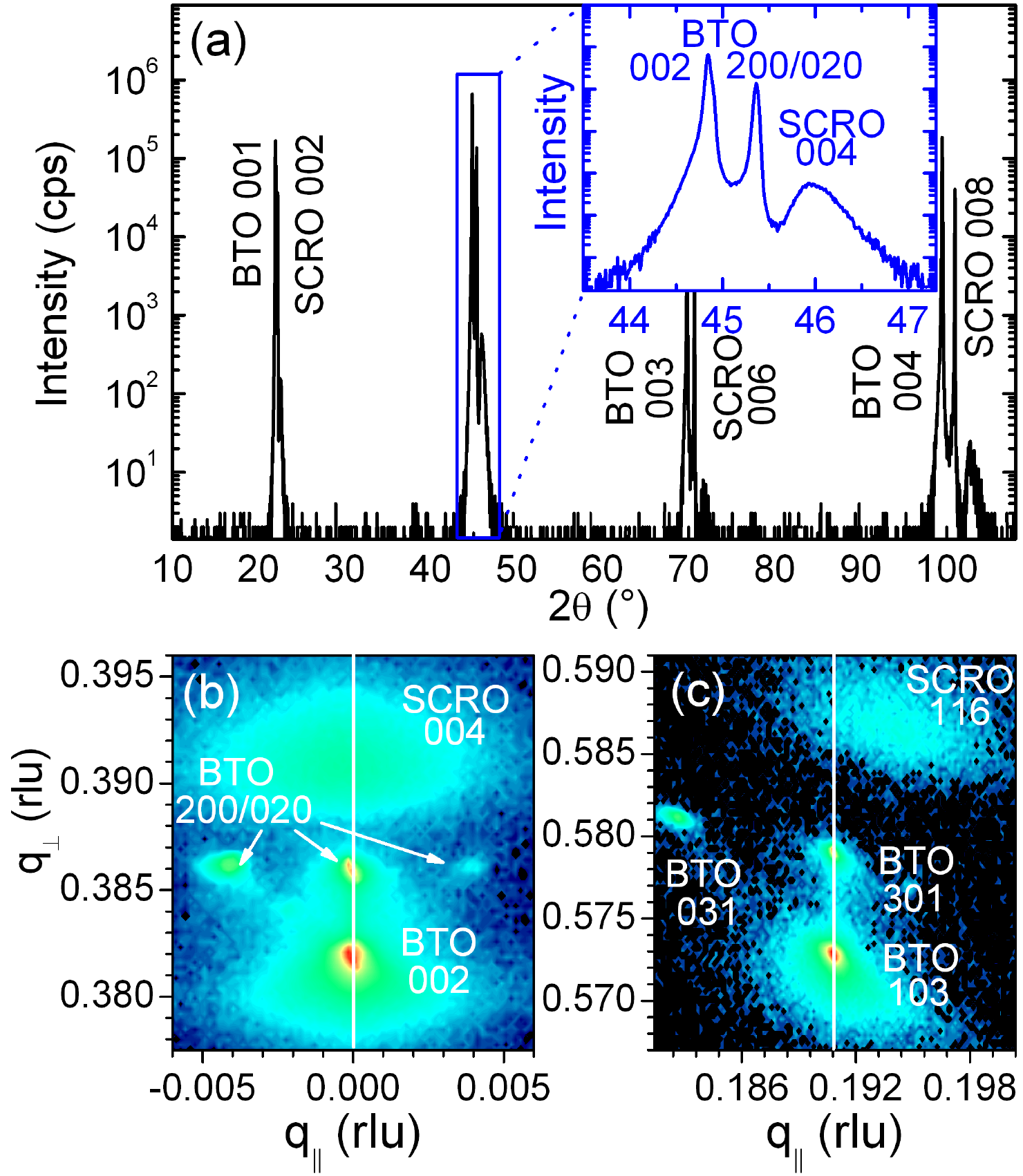}
    \caption{
HRXRD measurement of a 81\,nm SCRO film at 310\,K with the BTO
substrate in the tetragonal phase. (a) $\omega$-$2\theta$ scan. The inset
shows the SCRO (004) reflection and the BTO substrate reflection which is split into
$a$-domains (BTO (200)/(020)) and $c$-domains (BTO (002)). In the main figure
only the $c$-domains are indexed. (b,c) reciprocal space maps around the SCRO
(004) and (116) reflections. Clearly the different reflections of the BTO
domains are visible.}
    \label{Fig-XRD}
\end{figure}
%%%%%%%%%%%%%%%%%%%%%%%%%%%%%%%%%%%%%%%%%%%%%%%%%%%%%%%%%%%%%%%%%%%%%%

The structural properties of a 81\,nm SCRO film were studied by HRXRD
at 310\,K, in the tetragonal phase of BTO (Fig.~\ref{Fig-XRD}). The $\omega$-$2\theta$ scan (Fig.~\ref{Fig-XRD}(a)) reveals
no crystalline parasitic phases in the SCRO film. Moreover, the inset shows the
splitting of the BTO substrate reflection due to the presence of different
domains in the tetragonal phase ($a$-domains: BTO (200/020) and $c$-domains:
BTO (002)). A more detailed picture of the structural properties is obtained
from the reciprocal space maps (RSM) around the SCRO (004) and SCRO (116)
reflections (Fig.~\ref{Fig-XRD}(b,c)). Clearly no $q_\|$-shift of the
symmetric (004) SCRO reflection, but a distinct shift of the asymmetric (116)
SCRO reflection with respect to the corresponding BTO substrate reflections is
observed. This shows that the SCRO film grows $c$-axis oriented and is
partially relaxed with lattice parameters of $a_{\textrm{SCRO}}$=5.614\,\AA\
and $c_{\textrm{SCRO}}$=7.88\,\AA. Moreover, in Fig.~\ref{Fig-XRD}(b), tilted
$a$-domains with finite $q_\|$ values are visible. The two classes of $a$-domains
($a_{1}$: BTO (301) and $a_{2}$: BTO (031)) are clearly seen in
Fig.~\ref{Fig-XRD}(c). Considering the intensities of the reflections from both
$a$- and $c$-domains, an about equal amount of both domains is found for the
present substrate. This multi-domain state of the BTO substrate also results in
a considerable mosaic spread, as evident from the full width at half maximum
(FWHM) of 0.6$^\circ$ of the rocking curve of the SCRO (004) reflection, which
is about one order of magnitude larger than in SCRO films grown on STO
\cite {Gepraegs2009} and prevents resolving different
domains in the SCRO film. Additionally, the presence of a (101)
superstructure reflection (not shown) indicates an ordering of the Cr/Re
sub-lattice in our SCRO films. From the intensity ratio of the (101) and
(404) reflections, the amount of anti-site defects is estimated to be less than
30\%.

%%%%%%%%%%%%%%%%% FIGURE 2: Transport properties %%%%%%%%%%%%%%%%%%%%%%%%%%%%%%%%%%%%%
\begin{figure}[tb]
    \centering
    \includegraphics [width=0.85\columnwidth,clip=]{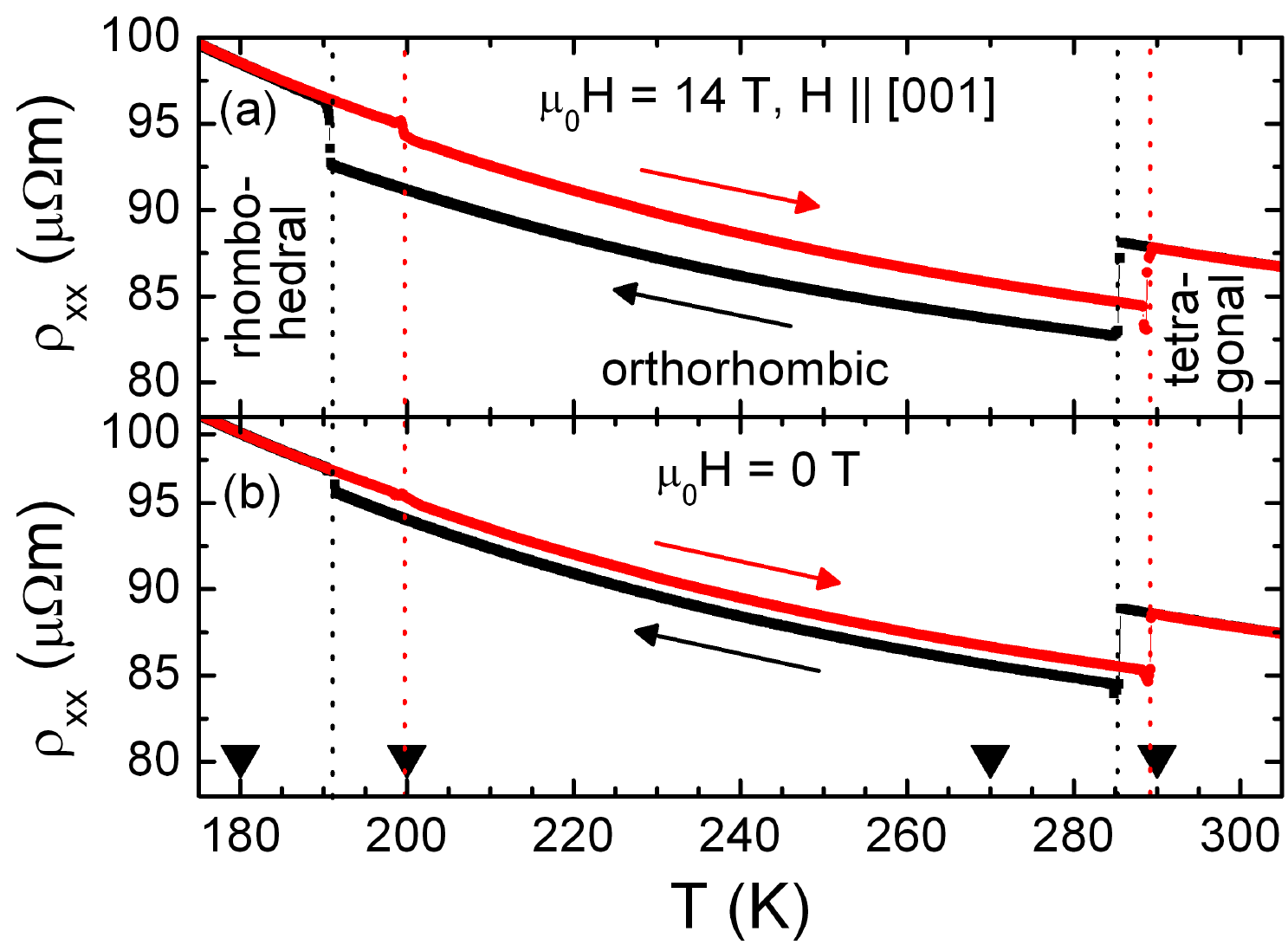}
    \caption{
Temperature dependence of the longitudinal resistivity $\rho_{xx}$ at (a)
$\mu_0H$=14\,T oriented along [001] and (b) 0\,T. Jumps
in the resistivity at the temperatures of the phase transitions (dotted lines)
of the BTO are observed. For the calculation of $\rho_{xx}$ a temperature
independent sample length and cross-sectional area was assumed.
The black triangles indicate the temperatures at which
the magnetization measurements in Fig. 3 were carried out.}
    \label{Fig-RT}
\end{figure}
%%%%%%%%%%%%%%%%%%%%%%%%%%%%%%%%%%%%%%%%%%%%%%%%%%%%%%%%%%%%%%%%%%%%%%

Whenever the BTO substrate crosses a structural phase transition, the
associated change of the lattice parameters should have a substantial effect on
the magnetic and electronic properties of the SCRO thin film clamped to its
surface \cite{Lee2000,Dale2003}. Fig.~\ref{Fig-RT} shows the temperature
dependent longitudinal resistivity $\rho_{xx}$ of a 70\,nm SCRO film,
patterned into a 375\,$\mu$m long and 50\,$\mu$m wide Hall bar. The curves in
Fig.~\ref{Fig-RT}(a) were measured in an external field of $\mu_0H$=14\,T,
orthogonal to the film plane, and the ones in Fig.~\ref{Fig-RT}(b) at
$\mu_0H$=0\,T. In both cases, a qualitatively similar $\rho_{xx}(T)$ is
observed: Upon cooling the sample from 300\,K (black lines), the resistivity
increases slightly until it suddenly drops at 285\,K, i.e.~at the temperature
at which the BTO crystal becomes orthorhombic. Between 285\,K and 191\,K, the
resistivity increases again continuously until it jumps to higher values at
191\,K, when the BTO transforms into the rhombohedral state. Further cooling
leads to a steadily increasing resistivity. The red curves in
Fig.~\ref{Fig-RT}, measured while warming up, reproduce the described behavior with a thermal hysteresis due to the first order phase transitions of
the BTO substrate. Please note that the difference in the resistivity for
warming and cooling is negligibly small in the rhombohedral and in the
tetragonal phase of the BTO substrate. Cracking of the BTO substrate thus is
not an issue. Furthermore, the resistivity in these two phases is also very
similar for the two magnetic fields $\mu_0H$=0\,T and 14\,T. Only the values in
the orthorhombic phase depend strongly on the sweep direction and the magnetic
field. Such a dependence was also observed in
La$_{0.5}$Sr$_{0.5}$MnO$_{3}$ thin films grown on BTO \cite{Dale2003}.
Chopdekar $et~al.$ attributed such a hysteresis to different ratios of $a$- and
$c$-domains in the orthorhombic phase depending on whether the previous phase
was rhombohedral or tetragonal \cite{Chopdekar2006}.

The resistivity jumps evident from Fig.~\ref{Fig-RT} are in the range of
several percent (up to 6.5\%). Discussing their origin one first has to
consider a simple geometry effect due to changes of the length $l$ and the
cross-sectional area $A$ of the current path assuming a constant resistivity
$\rho$. Since we did not control the ferroelectric domains in experiment, we
also considered a single domain substrate for simplicity. Furthermore, we assumed
that the SCRO film volume is conserved and that the film is fully strained.
Using the changes of the SCRO lattice constants at the BTO phase transitions \cite{Kay1949} we then estimate an upper limit for the expected
resistance changes of about 1\% for both transitions. This upper bound, as also
reported by Lee {\it et al.}\cite{Lee2000}, is much smaller than our observations
of up to 6.5\%. Therefore, a simple geometric effect to explain our data can be ruled out. Anisotropic magnetoresistance effects also
should be small, as the magnetization is essentially saturated at an external
field of 14\,T. On the other hand, it is well known that double perovskites
like SCRO are very sensitive towards distortions of the crystal and the
corresponding change in the overlap of the orbitals \cite{Philipp2003}. We therefore attribute the observed  resistance jumps to strain induced changes in the orbital configuration of SCRO.

%%%%%%%%%%%%%%%%% FIGURE 3: Magnetic properties %%%%%%%%%%%%%%%%%%%%%%%%%%%%%%%%%%%%%
\begin{figure}[tb]
    \centering
    \includegraphics [width=0.85\columnwidth,clip=]{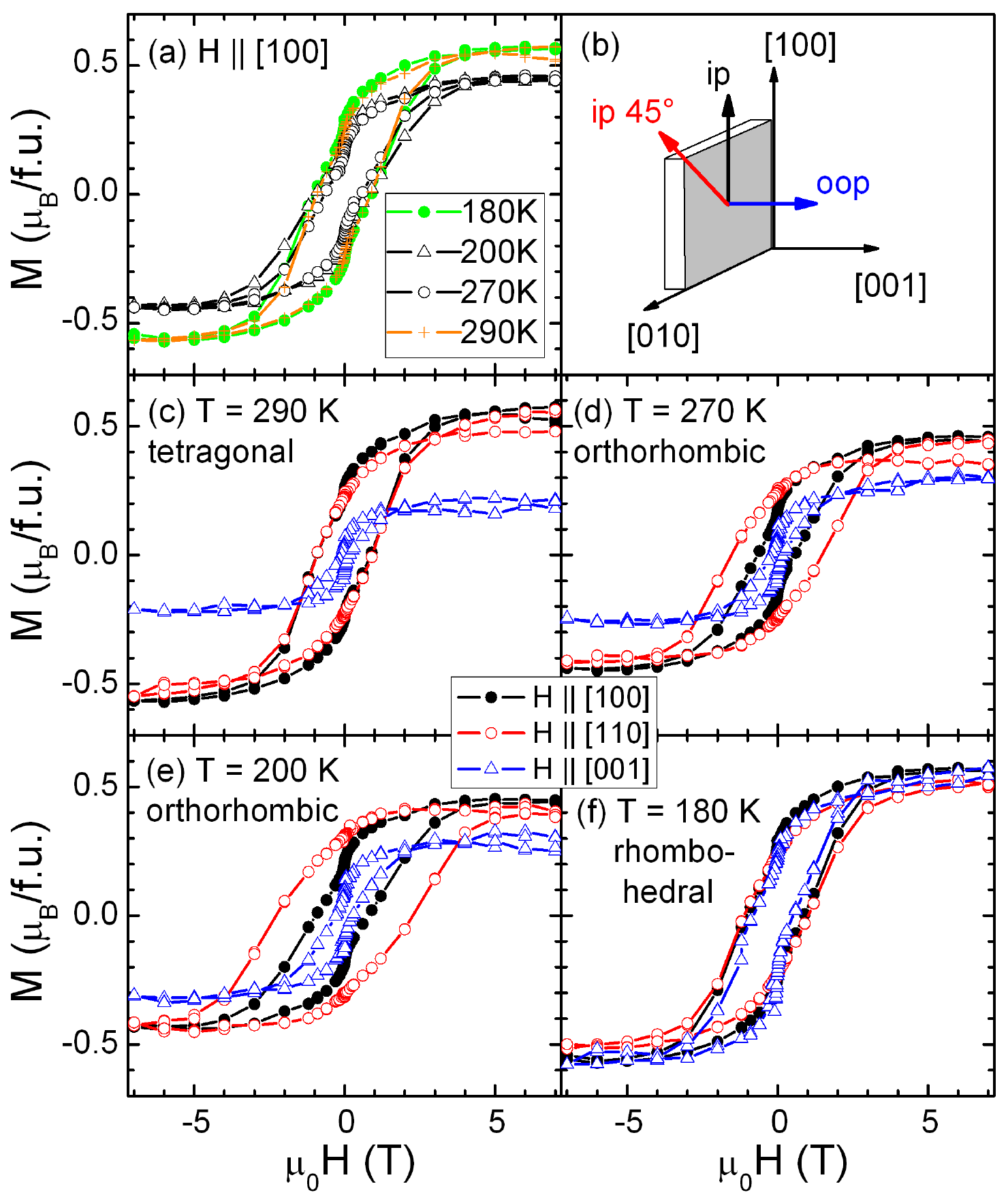}
    \caption{
Magnetization versus external magnetic field. (a) Hysteresis loops at
290\,K, 270\,K, 200\,K and 180\,K with the external magnetic field applied
along BTO [100]. (b) Illustration of the different orientations of the external
field. In panels (c)-(f) hysteresis loops at specific temperatures for
different orientations of the external magnetic field are shown.
        }
    \label{Fig-MH}
\end{figure}
%%%%%%%%%%%%%%%%%%%%%%%%%%%%%%%%%%%%%%%%%%%%%%%%%%%%%%%%%%%%%%%%%%%%%%

The various crystalline phases of the BTO substrate and the associated change
in the orbital configuration should also affect the magnetic properties of the
SCRO film. Fig.~\ref{Fig-MH}(a) shows the magnetization $M$ versus
external magnetic field $H$, oriented along BTO [100] for four different
temperatures: 290\,K, 270\,K, 200\,K, and 180\,K. The curves were obtained
on a single sample during one cooling cycle. All loops look very similar except
for the absolute value of the magnetization at high fields
(\textbar$\mu_0H$\textbar\textgreater4\,T). In the tetragonal (crosses) and the
rhombohedral phase (full circles) of BTO the $M$ values are nearly identical,
whereas in the orthorhombic phase (open circles and triangles) a reduction can
be observed. This indicates a reduced saturation magnetization which might be
due to the highly twinned BTO crystal in that phase.

In the following we investigate the field dependence of the magnetization in
the different BTO phases for three different orientations of the external
magnetic field: ip ($H \| [100]$), ip45 ($H \| [110]$) and oop ($H \| [001]$)
(see Fig.~\ref{Fig-MH}(b)). The temperatures of the field sweeps were chosen
slightly above and below the phase transition from tetragonal to orthorhombic
(290\,K and 270\,K) and from orthorhombic to rhombohedral (200\,K and 180\,K)
as indicated by the black triangles in Fig.~\ref{Fig-RT}. First we consider the
evolution of the in-plane anisotropy. In this case the external magnetic field
is applied in the film plane with two orientations: ip (full circles) and
ip45 (open circles). In the tetragonal phase of BTO, at 290\,K
(Fig.~\ref{Fig-MH}(c)), the two hysteresis loops are very similar. Both show a
coercive field of around 0.92\,T, and the magnetizations at
7\,T are nearly identical (0.6\,$\mu_{\rm{B}}$/f.u.). This suggests a
negligible in-plane magnetic anisotropy. In the orthorhombic phase of BTO, at
270\,K and 200\,K (Fig. 3(d,e)), the situation is completely different:
The coercivities of the ip and ip45 loops differ by more than 1\,T. In
particular, the hysteresis loops at 200\,K reveal a coercive field of
0.87\,T for the ip orientation of the external field and a much
larger value of 2.3\,T for the ip45 orientation. Thus a tremendous
in-plane magnetic anisotropy is present. Upon cooling into the rhombohedral
phase of BTO (Fig.~\ref{Fig-MH}(f)), the situation changes again: The coercive
fields and the magnetizations at 7\,T for both in-plane hystereses (ip and
ip45) as well as the oop loop (open triangles) are nearly identical. In other words, the
magnetic behavior appears isotropic, with no evidence of shape or crystalline
anisotropy contributions. This is remarkable, as it suggests that giant
strain-induced anisotropies of more than 1 Tesla are effective in the SCRO
film, compensating demagnetization.

In summary, we have shown that the strain associated with different crystalline
phases of a BaTiO$_3$ substrate induces pronounced changes both in the
resistivity and in the magnetic anisotropy of epitaxial Sr$_2$CrReO$_6$ thin
films. Abrupt jumps of up to 6.5\% in the resistivity, as well as
tremendous changes in the coercive field of more than $1.2$\,T  were observed
as a function of temperature. These observations can be consistently understood
considering orbital ordering and the strong electronic correlations in double
perovskite ferromagnets.

Financial support by the Deutsche Forschungsgemeinschaft via the priority
programs 1157 and 1285 (project nos. GR 1132/13 \& 1132/14), GO 944/3, and the
Excellence Cluster "Nanosystems Initiative Munich (NIM)" is gratefully
acknowledged.

\end{document}